\documentclass[conference]{IEEEtran}
\IEEEoverridecommandlockouts

\usepackage{cite}
\usepackage{amsmath,amssymb,amsfonts}

\usepackage{graphicx}
\usepackage{textcomp}
\usepackage{xcolor}
\usepackage{subcaption}
\usepackage{float}
\usepackage{setspace}

\usepackage[ruled,vlined,linesnumbered]{algorithm2e}

\def\BibTeX{{\rm B\kern-.05em{\sc i\kern-.025em b}\kern-.08em
    T\kern-.1667em\lower.7ex\hbox{E}\kern-.125emX}}
\begin{document}

\title{Edge-Assisted Accelerated Cooperative Sensing for
CAVs: Task Placement and Resource Allocation\\
}

\author{
    \IEEEauthorblockN{ Yuxuan Wang\IEEEauthorrefmark{1}\IEEEauthorrefmark{2}, Kaige Qu\IEEEauthorrefmark{3},
                      Wen Wu\IEEEauthorrefmark{1}, and Xuemin (Sherman) Shen\IEEEauthorrefmark{4}}
    \IEEEauthorblockA{\IEEEauthorrefmark{1}Frontier Research Center, Pengcheng Laboratory, China}
    \IEEEauthorblockA{\IEEEauthorrefmark{2}School of Electronic and Computer Engineering, Peking University, China}
    \IEEEauthorblockA{\IEEEauthorrefmark{3}School of Transportation Science and Engineering, Beihang University, China}
    \IEEEauthorblockA{\IEEEauthorrefmark{4}Department of Electrical and Computer Engineering, University of Waterloo, Canada}
    \IEEEauthorblockA{Email: ywang@stu.pku.edu.cn, kaigequ@buaa.edu.cn,  wuw02@pcl.ac.cn, sshen@uwaterloo.ca}

}

\maketitle

\begin{abstract}
In this paper, we propose a novel road side unit (RSU)-assisted cooperative sensing scheme for connected autonomous vehicles (CAVs), with the objective to reduce completion time of sensing tasks. Specifically, LiDAR sensing data of both RSU and CAVs are selectively fused to improve sensing accuracy, and computing resources therein are cooperatively utilized to process tasks in real time. To this end, for each task, we decide whether to compute it at the CAV or at the RSU and allocate resources accordingly. We first formulate a joint task placement and resource allocation problem for minimizing the total task completion time while satisfying sensing accuracy constraint. We then decouple the problem into two subproblems and propose a two-layer algorithm to solve them. The outer layer first makes task placement decision based on the Gibbs sampling theory, while the inner layer makes spectrum and computing resource allocation decisions via greedy-based and convex optimization subroutines, respectively. Simulation results based on the autonomous driving simulator CARLA demonstrate the effectiveness of the proposed scheme in reducing total task completion time, comparing to benchmark schemes.

\end{abstract}


\section{Introduction}
The recent advancement in environment sensing and vehicle-to-everything (V2X) communication technology abruptly shifts autonomous driving from a fiction to an exciting practice in the soon future. To realize autonomous driving, connected autonomous vehicles (CAVs) are equipped with advanced communication capabilities and diverse on-board sensors, such as light detection and ranging (LiDAR) sensors and cameras \cite{CAVsensor}. CAVs can bring a plethora of far-reaching and transformative benefits, including unprecedented user experiences, superior increased transportation efficiency, and tremendously improved road safety and air quality \cite{CAVdevelop}.

Leveraging V2X communication technology enables the sharing of sensing data among CAVs and roadside sensing devices, e.g., roadside LiDAR, to achieve \emph{cooperative sensing}, which can enhance sensing performance in terms of both expanded sensing range and improved sensing accuracy ~\cite{CPraw},~\cite{CPdecision}. 
Unlike standalone CAV sensing whose performance is easily affected by sensor quality and occlusions from obstacles, cooperative sensing fuses data from multiple LiDAR sensors which can effectively enhance the sensing accuracy ~\cite{rawlevelresearch}, ~\cite{cooperative}.

In the literature, several pioneering studies have been devoted to cooperative sensing in the context of CAVs. Ye \emph{et al}. proposed an accuracy-aware and resource-efficient cooperative sensing scheme, in which road side unit (RSU) and CAVs collaborate to assist an ego CAV to enhance sensing accuracy~\cite{dataquality}. Luo \emph{et al}. utilized sensing data from both CAVs and roadside LiDAR sensors, which are transmitted to the edge server for joint processing~\cite{rawlevelresearch2}. In the above works, sensing data is shared to enhance both resource utilization and sensing accuracy. However, RSU-assisted or edge-assisted cooperative sensing is seldom investigated. As a key component of vehicular networks, RSUs are equipped with advanced computing capabilities and LiDAR sensors with a wide view, which can assist CAVs to enhance the sensing accuracy \cite{RSUresource1},\cite{RSUsensor}.  

Designing an efficient edge-assisted cooperative sensing scheme encounters the following challenges. \emph{1) System heterogeneity} - CAVs in different locations have different regions of interest (RoI), and the data quality collected by different CAVs also varies due to their different views\cite{RoI}. The computing capabilities of the RSU and CAVs also differ from each other by more than tenfold \cite{RSUvehicleresource}. Such system heterogeneity poses significant influences on the effectiveness of cooperative sensing, with inappropriate cooperation schemes leading to increased task completion time and decreased sensing accuracy. \emph{2) Coupled decision making} - In edge-assisted cooperative sensing, the location where tasks are processed, i.e., task placement decision, and the amount of spectrum and computing resources that are allocated, i.e., resource allocation decisions are coupled, increasing the difficulty of problem-solving. Therefore, it is necessary to design a low-complexity algorithm to judiciously make task placement and resource allocation decisions.

In this paper, we first propose a novel cooperative sensing scheme for LiDAR-based CAV sensing. In the proposed scheme, the RSU assists each CAV in completing sensing tasks by providing computing capabilities and sensing data, while task placement and resource allocation decisions are made for those tasks. 
We adopt a model to characterize the relationship between fused data and accuracy.
Secondly, we formulate an optimization problem to minimize the total task completion time while satisfying the sensing accuracy constraint. 
The problem is a mixed-integer non-linear optimization problem.
To solve the problem, we decompose the optimization problem into task placement and resource allocation subproblems, and propose a two-layer algorithm. 
The outer layer makes the task placement decision using the Gibbs sampling method. 
The inner layer allocates optimal communication resources using a greedy-based algorithm and makes the optimal computing resource allocation decision through convex optimization methods. 
We use the autonomous driving simulator CARLA to simulate the scenario, generate point cloud data, and evaluate the performance.
The simulation results show that the proposed scheme effectively reduces the total task completion time while ensuring sensing accuracy.
The main contributions of this paper are summarized as follows. 
\begin{itemize}
\item We propose an edge-assisted cooperative sensing scheme for LiDAR-based sensing tasks of CAVs.
\item We formulate a joint task placement and resource allocation optimization problem to minimize the total task completion time.
\item We decouple the problem into two subproblems and then propose a two-layer algorithm to solve them, respectively.
\end{itemize}

\section{System Model}

\subsection{Considered Scenario}\label{AA}
As shown in Fig. 1, we consider an edge-assisted autonomous driving scenario on a bidirectional urban road segment covered by an RSU. The RSU is equipped with a LiDAR sensor and an edge server, providing both sensing and computing capabilities. The RSU stores AI model for object detection and classification for autonomous driving, e.g., SECOND \cite{SECOND}. The network controller at the RSU is in charge of decision-making. There are $K$ objects on the road within the coverage of the RSU. Let $\mathcal{K} = \{1, 2,\cdots, K\}$ denote the set of objects, where  $k \in \mathcal{K}$ represents the object index. Let $\mathcal{M} = \{1, 2,\cdots, M\}$ denote the set of CAVs, where  $m \in \mathcal{M}$ represents the CAV index. Each CAV also possesses sensing and computing capabilities and stores AI model the same AI model as the RSU. Let $\mathcal{M^+} = \{0, 1, \cdots, M\}$ represent all the nodes, including RSU and CAVs, where node 0 corresponds to the RSU.

\begin{figure}[t]
\centering
\includegraphics[width=\linewidth, trim={0cm 2.1cm 0cm 2.5cm}, clip]{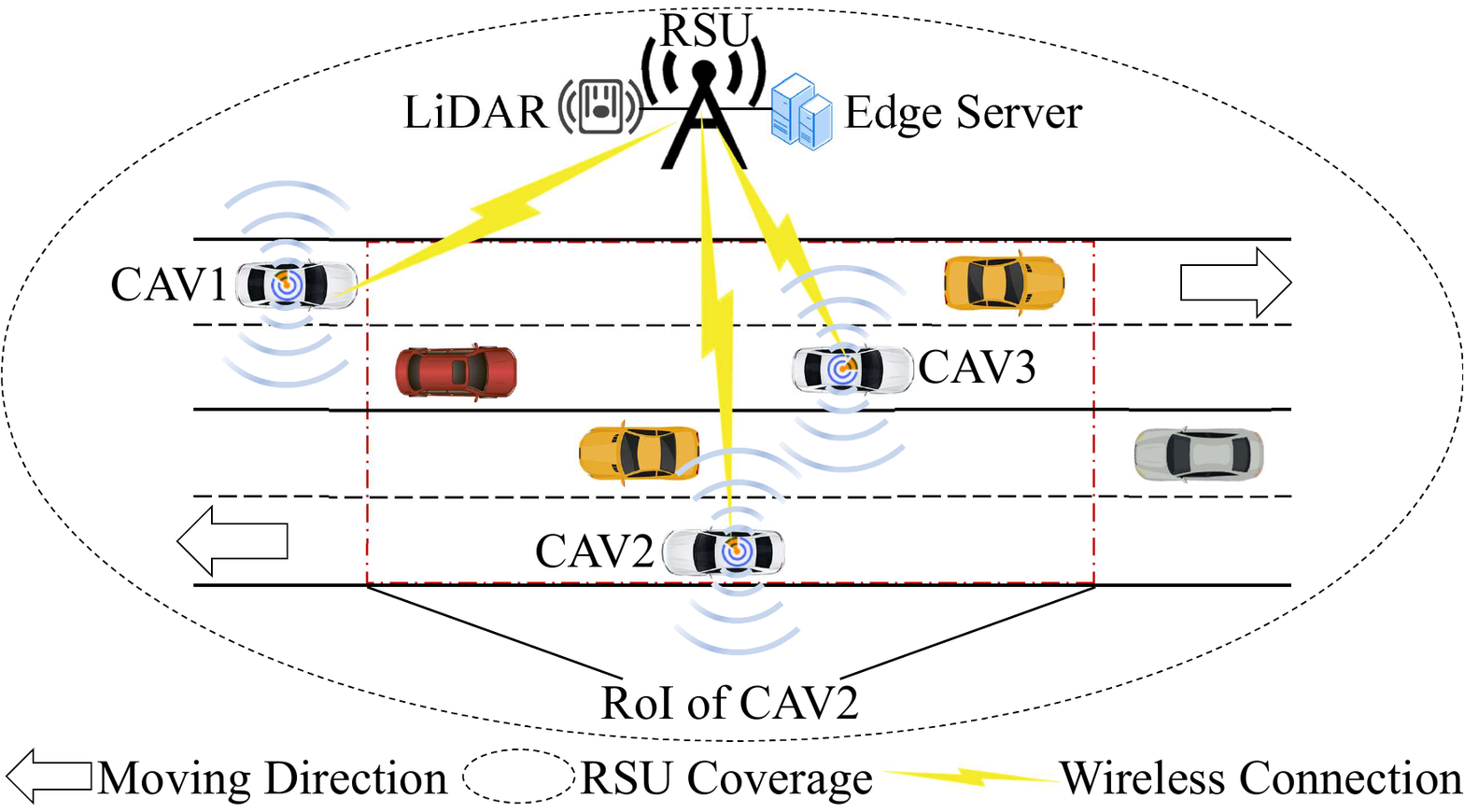}
\caption{Considered scenario.}
\label{fig1}
\end{figure}

Each CAV has its own unique RoI and the CAVs sense objects within their respective RoIs. For each CAV, its RoI is defined as a rectangle, whose length extends $R_1$ meters behind the CAV and $R_2$ meters ahead, and the width $L$ equals the width of the road. As shown in Fig. \ref{fig1}, there are three white CAVs from left to right, namely CAV 1, CAV 2, and CAV 3. The RoI of CAV 2 is shown in the red dotted box. Let $\mathcal{U} = \{u_{k,m},\ \forall k \in \mathcal{K}, m \in \mathcal{M}\}$ denote a binary matrix, where $u_{k,m} = 1$ indicates that object $k$ is within the RoI of CAV $m$, and $u_{k,m} = 0$ otherwise. With the assistance of the RSU, detection and classification for each object can be performed independently. Therefore, an object that a CAV needs to detect and classify is a sensing task, which can be processed in parallel. We let $\phi_{k,m}$  represent the task of the object $k$ for the $m$-th CAV. Let $\mathcal{M}^k$ denote the set of CAVs that require processing task $k$, represented by $\mathcal{M}^k = \{m  \in \mathcal{M}|u_{k,m}=1\}$.

We propose an edge-assisted cooperative sensing scheme, leveraging the RSU to enhance sensing accuracy and reduce CAV task completion time. The process consists of the following four stages:
\begin{enumerate}
    \item RSU broadcast: The RSU performs bounding box detection for the objects within its coverage and then broadcasts the bounding box parameters and data within each object to each CAV.
    
    \item CAV demand upload: According to the information broadcast by RSU, each CAV determines the objects within its RoI, that is, the number of tasks that need to be processed. 
Some of these objects may be obscured and not detected by the CAV. 
Each CAV uploads its task requirement, computing resource, and data quality information to RSU.
    
    \item Control module decision-making: The control module on RSU makes decisions considering the demands of the CAVs and the available resources, then distributes decisions accordingly.

    \item Sensing task processing: For tasks processed locally, sensing data from the CAV and the RSU fused and processed. For tasks processed at the RSU, the RSU waits to receive all necessary data from the CAVs and sends the results back once processing is complete.
\end{enumerate}

\subsection{Sensing Data Model}

The LiDAR sensor of node $m$ generates 3D point cloud data, represented as the set ${\mathcal{N}}_m = \{(x^i_m,y^i_m,z^i_m)  \mid i = 1, 2, \dots, D_m\}$, where $D_m$ denotes the total number of points in ${\mathcal{N}}_m$. All point cloud data are aligned through global coordinate transformation\cite{RSUresource2}.

The spatial coordinates of an object $k \in \mathcal{K}$ are defined by a 3D cuboid bounding box that encloses the object. This bounding box is expressed as a 6-tuple, $\mathcal{C}_k = (x_k,y_k,z_k,l_k^x,l_k^y,l_k^z)$, where $x_k,y_k,z_k$ denote the cuboid center’s 3D coordinates and $l_k^x,l_k^y,l_k^z$ denote the cuboid’s dimensions along the $x$, $y$, and $z$ axes. 
The sensing data for object $k$ at node $m$, termed object data, is denoted by ${\mathcal{N}}_{k,m}$, which can be extracted from the data ${\mathcal{N}}_m$ using the bounding box parameter $\mathcal{C}_k$.

We use the data quality $\boldsymbol{q}_{k,m} \in \mathbb{R}^{1 \times J^3}$ \cite{dataquality} to describe the number of points and spatial distribution of the data ${\mathcal{N}}_{k,m}$. Each object's bounding box is evenly divided into $J$ segments along each axis, resulting in $J^3$ sub-boxes. The number of points within each sub-box is then counted to represent the data quality $\boldsymbol{q}_{k,m}$ of ${\mathcal{N}}_{k,m}$.

\subsection{Sensing Task Model}
The RSU senses all objects within its coverage area, while each CAV senses the objects within its own RoI. Through low-resolution data fusion, the RSU can perform object bounding box detection within its coverage area. We ignore the resource cost of bounding box detection and assume that all object bounding boxes are successfully detected without loss of generality. Since object classification requires processing high-resolution point cloud data, we primarily focus on the object classification phase.

The processing location needs to be decided for each object $k$ for each CAV $m$.
 
Let $\boldsymbol{E} = \{e_{k,m},\ \forall k \in \mathcal{K}, m \in \mathcal{M}^k\}$  denote the binary task placement decision, which decision variable $e_{k,m} = 0$ indicating that task $\phi_{k,m}$ is placed at CAV locally, and $e_{k,m} = 1$ indicating that task $\phi_{k,m}$ is placed at RSU. For the same object $k$, if multiple CAVs need to send their sensing tasks for this object to the RSU for processing, we refer to these tasks as the corresponding tasks of the object.

Different task processing locations determine different data fusions. If CAV $m$ processes task $\phi_{k,m}$ locally, the CAV fuses its sensing data with object data received from the RSU. The fused data, which needs to be processed at the CAV, is represented by $\mathcal{N}_{k,m}^{'}= u_{k,m}\{(1-e_{k,m})\mathcal{N}_{k,m} \cup \mathcal{N}_{k,0}\}$. 
Let $\mathcal{M}^k_{RSU}$ denote the set of CAVs that process the object $k$ at the RSU, represented by $\mathcal{M}^k_{RSU} = \{m  \in \mathcal{M}|u_{k,m}=1,e_{k,m}=1\}$. If multiple CAVs require processing the same object $k$ at the RSU, the RSU fuses the received object data from $\mathcal{M}^k_{RSU}$ with its own data and distributes the results to the required CAVs. The fused data, which needs to be processed at the RSU, is represented by $\mathcal{N}_{k,0}^{'}= u_{k,m}\{( \cup_{m \in \mathcal{M}^k_{RSU}} \mathcal{N}_{k,m}) \cup \mathcal{N}_{k,0}\}$. The quality of the fused data $\boldsymbol{q}^{\prime}_{k,m}$ can be calculated using the method mentioned in Section II-B.

\subsection{Communication Model}
If the task $\phi_{k,m}$ is placed at the RSU for processing, i.e., $e_{k,m} = 1$, CAV $m$ needs to transmit data $\mathcal{N}_{k,m}$ to the RSU.
Let $\xi$ represent the data size (in bits) per observation point, the total size of the sensing data transmitted from CAV $m$ to RSU is denoted by $S_{k,m} = \xi |\mathcal{N}_{k,m}|,\forall k \in \mathcal{K}, m \in \mathcal{M}^k_{RSU}$.

With the consideration that the total bandwidth $B$ is available, subcarrier allocation is performed for each task $\phi_{k,m}$ that needs to transmit data. Let $\boldsymbol{B} = \{b_{k,m},\  \forall k \in \mathcal{K}, m \in \mathcal{M}^k_{RSU}\}$ represent allocation decision, where $b_{k,m}$ represents the number of subcarriers. Each subcarrier has a bandwidth of $B_s$, the bandwidth allocation variable should satisfy the following constraint
\begin{equation}
\label{eq:constraint5}
0 \leq b_{k,m} \leq \frac{B}{B_s}\ \ \ \ \  \forall k \in \mathcal{K}, m \in \mathcal{M}^k_{RSU}.
\end{equation}

The sum of the allocated bandwidth should equal the total bandwidth $B$, given by
\begin{equation}
\label{eq:constraint2}
 \sum_{ k \in \mathcal{K}} \sum_{m \in {M}^k_{RSU}}\, b_{k,m}B_s = B.
\end{equation}

Then, the transmission time between CAV $m$ and RSU for task $\phi_{k,m}$ is given by $t_{km,R} = S_{k,m}/R_{m}$. Let $R_{m}$ denote the transmission rate between CAV $m$ and the RSU, which can be calculated using Shannon's formula. For object $k$, if multiple CAVs process task $\phi_{k,m}$ at the RSU, the RSU will process the task after receiving data from all these CAVs. Therefore, the transmission time for task $\phi_{k,m}$ is given by $t_{k,m}^{trans} = \max_{m\in\mathcal{M}^{k}_{RSU}}\{t_{km,R}\}$. The time of RSU data broadcasting is denoted as $t_{broad}$. Since the size of processing results is small, we ignore the time it takes for RSU to send results. Then, the total communication time for task $\phi_{k,m}$ is denoted as $
t_{k,m}^{com} = u_{k,m}t_{broad}+t_{k,m}^{trans},\forall k \in \mathcal{K}, m \in \mathcal{M}^k_{RSU}$.

\subsection{Computing Model}

The computing demand of a task is proportional to the number of points in the point cloud that need to be processed. Let $\omega$ denote the computation intensity (in cycles/point) representing the average CPU cycles for computing one point of the sensing data. 
For object $k$, let $\mathcal{M}^k_{comp} = \{m \in \mathcal{M^{+}} | \mathcal{N}_{k,m}^{'} \neq \varnothing\}$ be the set of nodes that need to process it.
The computing demand of object $k$ node $m$ is denoted by $
C_{k,m} = \omega |\mathcal{N}_{k,m}^{'}|, \forall k \in \mathcal{K}, m \in \mathcal{M}^k_{comp}$.

Let $\boldsymbol{V} = \{v_{k,m},\forall k \in \mathcal{K}, m \in \mathcal{M}^k_{comp}\}$ be a continuous computing resource allocation decision, with $v_{k,m}$ indicating the proportion of computing resources used at computing node $m$ for object $k$.
The computing resource allocation variable should satisfy the following conditions
\begin{equation}
\label{eq:constraint6}
0 \leq v_{k,m} \leq 1 \ \ \ \ \ \ \ \ \forall k \in \mathcal{K}, m \in \mathcal{M}^k_{comp}.
\end{equation}

The sum of the allocated computing resource fractions should not exceed 1, given by
\begin{equation}
\label{eq:constraint3}
 \sum_{k \in \mathcal{K}} v_{k,m} \leq 1  \ \ \ \ \ \  \forall m \in \mathcal{M^+}.
\end{equation}
Assume that RSU starts processing tasks immediately after receiving the required data. 
For task $\phi_{k,m}$, if the CAV $m$ offloads it to the RSU, the total computation time is given by $
t_{k,m}^{comp} = {C_{k,0}}/{v_{k,0}f_{0}}, \forall k \in \mathcal{K}, m \in \mathcal{M}^{k}_{RSU}$. For task $\phi_{k,m}$, if the CAV $m$ performs local computing, the total computation time is given by $
t_{k,m}^{comp} = {C_{k,m}}/{{v_{k,m}}f_{m}}, \forall k \in \mathcal{K}, m \in \mathcal{M}^k\setminus\mathcal{M}^{k}_{RSU}$,where ~$f_{m}$ represents the computing capability of node $m$.

\subsection{Sensing Accuracy Model}\label{SCM}
The result of the classification task is a multi-dimensional estimated class probability vector, where a higher estimated probability indicates greater confidence from the model that an object is present and belongs to this class. Therefore, the estimated probability of the true class is used as the accuracy metric. We consider that the fused data $\mathcal{N}_{k,m}^{'}$ is processed by the AI model, and the accuracy $a_{k,m}$ of the model's output, which depends on the fused data quality $\boldsymbol{q}^{\prime}_{k,m}$, can be evaluated using an accuracy estimation function. 

To facilitate better decision-making, we use a deep neural network (DNN) model to fit the relationship between accuracy $a_{k,m}$ and data quality $\boldsymbol{q}^{\prime}_{k,m}$. The model's input includes the data quality and the size of the bounding box, and the output is the accuracy. It can be represented as $a_{k,m} = f(\boldsymbol{q}^{\prime}_{k,m},l_k^x,l_k^y,l_k^z)$, where $\boldsymbol{q}^{\prime}_{k,m}$ represents the data quality, $l_k^x,l_k^y,l_k^z$ represents the bounding box size, and $f$ is the function learned by the model to estimate the accuracy $a_{k,m}$\cite{dataquality}. We vary different scenarios and topologies to collect a large amount of data regarding data quality and accuracy for the offline pre-training of the model. The trained model is then deployed on RSU to assist in decision-making.

\section{Problem Formulation}

Assume that all CAVs upload information immediately after determining the object in their RoI. The total task completion time for any object at any CAV in $\mathcal{M}$ is denoted by $t_{k,m} = t_{k,m}^{comp}+t_{k,m}^{com}, \forall k \in \mathcal{K}, m \in {\mathcal{M}^k}$.
The task accuracy requirement is that the average accuracy of all objects for each CAV should exceed a threshold. To satisfy the accuracy requirement of each CAV, we establish the following constraint:

\begin{equation}
\label{eq:constraint4}
\frac{\sum_{k \in \mathcal{K}} a_{k,m}u_{k,m}}{\sum_{k \in \mathcal{K}} u_{k,m}}   \geq A  \ \ \ \ \ \ \ \ \ \ \ \ \forall m \in \mathcal{M},
\end{equation}
where $A$ represents the required minimum sensing accuracy.

The objective is to minimize the total completion time of all tasks while ensuring accuracy. We formulate a joint task placement, communication resource allocation, and computing resource allocation problem as an optimization problem, i.e.,
\begin{subequations}
    \label{P0}
    \begin{align}
        \mathbf{P}0:&\min_{\boldsymbol{E}, \boldsymbol{B}, \boldsymbol{V}}  \sum_{k \in \mathcal{K}} \sum_{m \in {\mathcal{M}^k}} t_{k,m}\notag \\
        &\text{s.t.}\  \eqref{eq:constraint5},\ \eqref{eq:constraint2},\ \eqref{eq:constraint6},\ \eqref{eq:constraint3},\ \eqref{eq:constraint4}, \notag \\
        & e_{k,m} \in \{0,1\},\ \ \   \forall k \in \mathcal{K}, m \in \mathcal{M}^k, \tag{\ref{P0}a} \label{P0a} \\
        & b_{k,m} \in \mathbb{Z},\ \ \   \forall k \in \mathcal{K}, m \in \mathcal{M}^k_{RSU}, \tag{\ref{P0}b} \label{P0b} \\
        & 0 \leq v_{k,m} \leq 1,\ \ \  \forall k \in \mathcal{K}, m \in \mathcal{M}^k_{comp}.\tag{\ref{P0}c} \label{P0c}
    \end{align}
\end{subequations}

Problem $\mathbf{P}0$ has constraints in terms of task placement, bandwidth allocation, computing resource allocation, and accuracy. Constraint \eqref{P0a}  is the binary constraint on the task placement variable. Constraints \eqref{eq:constraint5}, \eqref{eq:constraint2}, and \eqref{P0b} are bandwidth allocation constraints that ensure the bandwidth allocation decision is feasible and the allocated bandwidth is equal to the total available bandwidth. Constraints \eqref{eq:constraint6}, \eqref{eq:constraint3}, and \eqref{P0c} are computing resource allocation constraints, confirming the feasibility of the decision and ensuring that the resources assigned to tasks at each computing node do not exceed the total available capacity of the node. Constraint \eqref{eq:constraint4} guarantees that the accuracy of all CAVs exceeds the threshold.

Problem $\mathbf{P}0$ is a mixed-integer non-linear optimization problem. Solving this problem using traditional optimization methods is challenging. We divide Problem $\mathbf{P}0$ into two sub-problems: i) communication and computing resource allocation subproblem, and ii) task placement subproblem. We design a two-layer algorithm to solve the problem, with the outer layer obtaining $\boldsymbol{E}$ and the inner layer obtaining $\boldsymbol{B},\boldsymbol{V}$.

\section{Proposed Solution}

\subsection{Communication and Computing Resource Allocation}
\begin{algorithm}[t]
\small
\SetAlgoLined
\caption{Greedy-Based Subcarrier Allocation Algorithm}
\label{algo1}

Initialization: $b_{k,m} = 1$ , $\forall k \in \mathcal{K}$, $m \in \mathcal{M}^k_{RSU}$\;

\For{iteration $= 1, 2, \dots, \frac{B}{B_s} - |\boldsymbol{B}|$}{
    Calculate transmission time $\Omega$ based on the objective function of $\mathbf{P}2$;

    \For{each $k \in \mathcal{K}$,$m \in \mathcal{M}^k_{\text{RSU}}$}{
        Increment allocation: $\hat{b}_{k,m} = b_{k,m} + 1$\;
        Update allocation set $\hat{\boldsymbol{B}}$: only update $\hat{b}_{k,m}$\;
        Calculate new transmission time: $\Omega_{new}$ based on the objective function of $\mathbf{P}2$\;
    }
    Select $(k^\star, m^\star)$ that maximizes the reduction in transmission time: $(k^\star, m^\star) = \arg\max_{k \in \mathcal{K},m \in \mathcal{M}^k_{RSU}} \{\Omega - \Omega_{new}\}$\;
    Allocate an additional subcarrier to $k^\star, m^\star$: $b_{k^\star, m^\star} = b_{k^\star, m^\star} + 1$\;
}
\end{algorithm}

The RSU starts processing object $k$ only after all required CAV data is collected. Therefore, the transmission time corresponding to a task is influenced by the communication resource allocation. In addition, the task can be placed in the CAV for local computing or in the RSU for computing. It is possible that multiple tasks need to be processed locally in the CAV or at the RSU, so the computing resource allocation for each task will affect the overall computation time. Therefore, we study communication resource allocation and computing resource allocation to minimize the sum of transmission time and computation time given the task placement decision. In contrast, in $t_{k,m}^{com}$, only $t_{k,m}^{trans}$ is affected by $\boldsymbol{B}$, while $u_{k,m}t_{broad}$ is not related to $\boldsymbol{B}$. Based on $\mathbf{P}0$, the problem of minimizing the sum of transmission time and computation time can be formulated as follows
\begin{equation}
\label{P1}
\begin{aligned}
\mathbf{P}1:\ &\min_{\boldsymbol{B},\boldsymbol{V}} \sum_{ k \in \mathcal{K}}\sum_{m \in {\mathcal{M}^k}} (t_{k,m}^{trans} + t_{k,m}^{comp})\\
&\text{s.t.}\ \eqref{eq:constraint5},\ \eqref{eq:constraint2},\ \eqref{eq:constraint6},\ \eqref{eq:constraint3},\ \eqref{P0b}, \text{and} \ \eqref{P0c}.\\
\end{aligned}
\end{equation}

Problem $\mathbf{P}1$ is a mixed-integer programming problem that includes an integer variable $\boldsymbol{B}$ and a continuous variable $\boldsymbol{V}$. Since communication resource allocation does not influence the total computation time, and the allocation of computing resources does not impact the transmission time, $\mathbf{P}1$ can be further divided into two subproblems $\mathbf{P}2$ and $\mathbf{P}3$.

Problem $\mathbf{P}2$ represents the allocation subproblem of communication resources, which aims to determine the allocation of communication resources to minimize the transmission time:
\begin{equation}
\label{P2}
\begin{aligned}
\mathbf{P}2:\ &\min_{\boldsymbol{B}} \sum_{ k \in \mathcal{K}}\sum_{m \in \mathcal{M}^k_{RSU}}t_{k,m}^{trans}\\
&\text{s.t.}\ \eqref{eq:constraint5},\ \eqref{eq:constraint2},\text{and} \ \eqref{P0b}.\\
\end{aligned}
\end{equation}

To address this problem, we propose a greedy-based subcarrier allocation algorithm. First, each transmission task of every CAV is assigned a subcarrier, forming an initial decision. Then, each subcarrier is allocated to the transmission task that can minimize the overall transmission time to the greatest extent, until all subcarriers are fully allocated, as described in Alg. \ref{algo1}.

Problem $\mathbf{P}3$ represents the computing resources allocation subproblem, the goal of which is to determine the allocation of computing resources to minimize the total computation time:
\begin{equation}
\label{P3}
\begin{aligned}
\mathbf{P}3:\ &\min_{\boldsymbol{V}}\sum_{ k \in \mathcal{K}}\sum_{m \in \mathcal{M}^{k}}t_{k,m}^{comp}\\
&\text{s.t.}\ \eqref{eq:constraint6},\ \eqref{eq:constraint3},\text{and} \ \eqref{P0c}.\\
\end{aligned}
\end{equation}

The computation time based on $C$ and $\boldsymbol{V}$ represents the computation time for a single node and a single object. However, for an object processed on the RSU, multiple CAVs may offload the corresponding tasks to the RSU for computing. In this case, the time required to process the object on the RSU is effectively the computation time for multiple CAVs. Since our optimization objective is to minimize the time for all CAVs across all tasks that need to be processed, we need to calculate the time taken by each CAV to complete each task. Therefore, we introduce a time factor $\eta$, where $\eta_{k,m}$ represents the number of CAVs that depend on the processing time of the $k$ object on the $m$ computing node. When $m>0$, it refers to a CAV computing node, and $\eta_{k,m} = u_{k,m}(1-e_{k,m})$. When $m=0$, it refers to the RSU computing node, and $\eta_{k,0} = len(\mathcal{M}^k_{RSU})$. So, ${\sum_{ k \in \mathcal{K}}\sum_{m \in \mathcal{M}^k}t_{k,m}^{comp}}$ can be rewritten as${\sum_{ k \in \mathcal{K}}\sum_{m \in \mathcal{M}^k_{comp}}\frac{\eta_{k,m}C_{k,m}}{{v_{k,m}}f_{m}}}$.

We derive the closed-form solution for Problem $\mathbf{P}3$ through the following steps. First, it is established that the rewritten Problem $\mathbf{P}3$ qualifies as a convex optimization problem. The objective function is ${\sum_{ k \in \mathcal{K}}\sum_{m \in \mathcal{M}^+}\frac{\eta_{k,m}C_{k,m}}{{v_{k,m}}f_{m}}}$. The second-order derivative of the objective function shows $\frac{2\eta_{k,m}C_{k,m}}{{v_{k,m}^3}f_{m}} > 0$. Furthermore, the inequality constraint is linear, which implies that the problem qualifies as a convex optimization problem. Second, a Lagrange function is formulated for the problem without taking the inequality constraints into account, $\mathcal{L}( \boldsymbol{V},\lambda) = \sum_{ k \in \mathcal{K}}\sum_{m \in \mathcal{M}^k_{comp}}\frac{\eta_{k,m}C_{k,m}}{{v_{k,m}}f_{m}} + \lambda(\sum_{ m \in \mathcal{M}^k_{comp}} v_{k,m} -1)$, where $\lambda$ denotes the Lagrange multiplier. Based on Karush–Kuhn–Tucker conditions, we can obtain $\frac{ \partial {\mathcal{L}( \boldsymbol{V},\lambda)}}{\partial v_{k,m}} = -\frac{\eta_{k,m}C_{k,m}}{{v_{k,m}^2}f_{m}} + \lambda = 0, \forall k \in \mathcal{K}, m \in \mathcal{M}^k_{comp}$. By solving this equation, we can obtain $v_{k,m}^{\star} = \sqrt{\eta_{k,m}C_{k,m}}/ \lambda f_{m}$. The complementary slackness condition can be written as $\sum_{ k \in \mathcal{K}}{v_{k,m}^{\star}}-1=0 $. Substituting $v_{k,m}^{\star}$ into the complementary slackness condition,  the optimal value of $\lambda$ is given by $\lambda^{\star} = {(\sum_{ k \in \mathcal{K}} \sqrt {\eta_{k,m}C_{k,m})}}^2/f_{m} $. Based on the previously discussed equation, it is evident that $\lambda$ assumes a positive value, which consequently implies that the set ${v_{k,m}^{\star}}$ consists of positive values as well. This automatically fulfills the constraint \eqref{P0c}, i.e., ${v_{k,m} }\geq 0$. The optimal computing resource allocation for problem $\mathbf{P}3$ is given by
\begin{equation}
\label{vsolution}
v_{k,m}^{\star} = \frac{\sqrt {\eta_{k,m}C_{k,m}}}{\sum_{ k \in \mathcal{K}}\sqrt {\eta_{k,m}C_{k,m}}} \ \ \ \ \ \forall k \in \mathcal{K}, m \in \mathcal{M}^k_{comp}.
\end{equation}
\begin{algorithm}[t]
\small
\SetAlgoLined
\caption{Two-Layer Joint Task Placement and Resource Allocation Algorithm}
\label{algo2}

Initialization: Randomly generate a feasible decision $\boldsymbol{E}$ and calculate $\boldsymbol{B}$, $\boldsymbol{V}$ based on Alg. 1 and Eq. \eqref{vsolution}, then calculate the objective value $\Delta$ \;

\For{iteration $= 1, 2, \dots, T$}{

    Generate a new candidate decision $\hat{\boldsymbol{E}}$ by changing a single value in $\boldsymbol{E}$\;
    
    \While{$\hat{\boldsymbol{E}}$ is infeasible}{
        Randomly change a single value in $\hat{\boldsymbol{E}}$\;
        Check feasibility of $\hat{\boldsymbol{E}}$ using Constraint \eqref{eq:constraint4} \;
    }
    
    Calculate $\hat{\boldsymbol{B}}$ and $\hat{\boldsymbol{V}}$ based on Alg. 1 and Eq. \eqref{vsolution}\;
    Calculate $\hat{\Delta}$ for $\hat{\boldsymbol{E}}$ based on $\hat{\boldsymbol{B}}$ and $\hat{\boldsymbol{V}}$ \hspace{0pt}\;
    $\gamma \gets \frac{1}{1 + \exp((\hat{\Delta} - \Delta) / \tau)}$\;
    With probability $\gamma$, set $\boldsymbol{E} \gets \hat{\boldsymbol{E}}$, $\boldsymbol{B} \gets \hat{\boldsymbol{B}}$, $\boldsymbol{V} \gets \hat{\boldsymbol{V}}$, $\Delta \gets \hat{\Delta}$\;
    With probability $(1 - \gamma)$, keep $\boldsymbol{E}$, $\boldsymbol{B}$, $\boldsymbol{V}$ unchanged\;

}

Return $\boldsymbol{E}$, $\boldsymbol{B}$, and $\boldsymbol{V}$, when the stopping criterion is satisfied; Otherwise, go to Line 2.
\end{algorithm}

\setlength{\belowcaptionskip}{-2mm}

\begin{figure*}[t] 
    \centering
    \begin{subfigure}{0.3\textwidth}
        \includegraphics[width=\textwidth,trim={0mm 8mm 0mm 8mm}, clip]{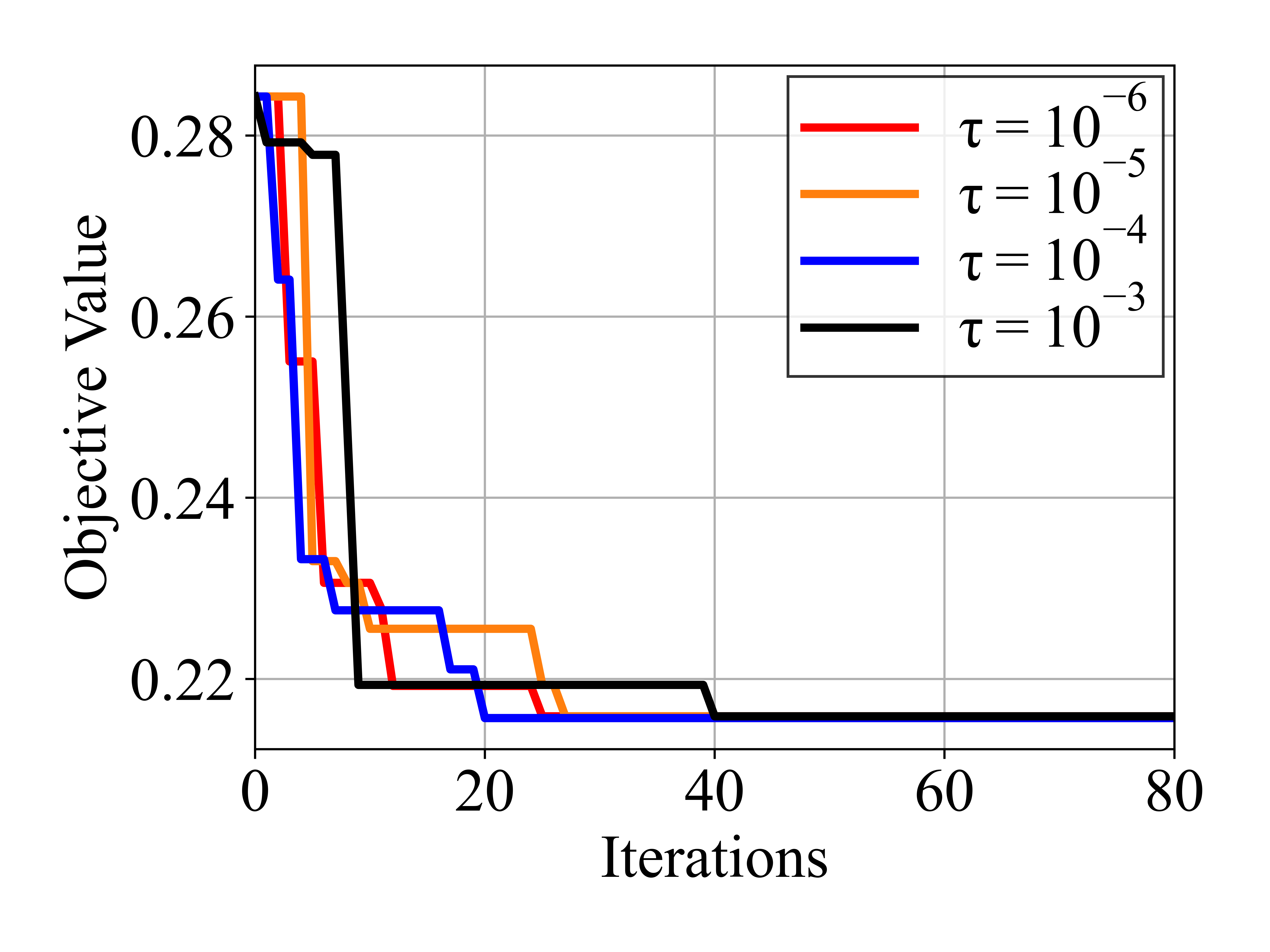}
        \caption{Convergence performance}
        \label{fig:image1}
    \end{subfigure}%
    \begin{subfigure}{0.3\textwidth}
        \includegraphics[width=\textwidth,trim={0mm 8mm 0mm 8mm}, clip]{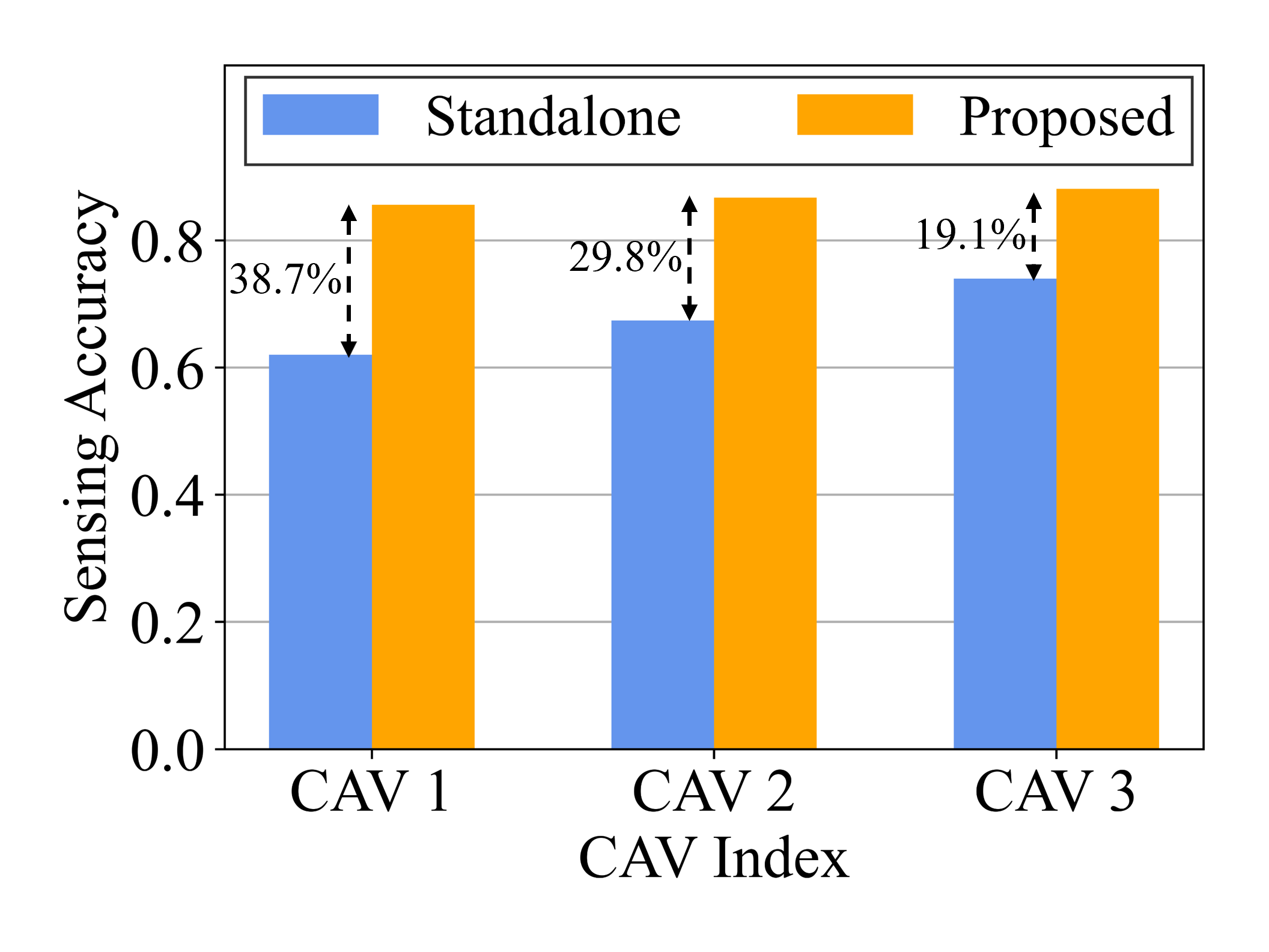}
        \caption{Sensing accuracy}
        \label{fig:image2}
    \end{subfigure}%
    \begin{subfigure}{0.3\textwidth}
        \includegraphics[width=\textwidth,trim={0mm 8mm 0mm 8mm}, clip]{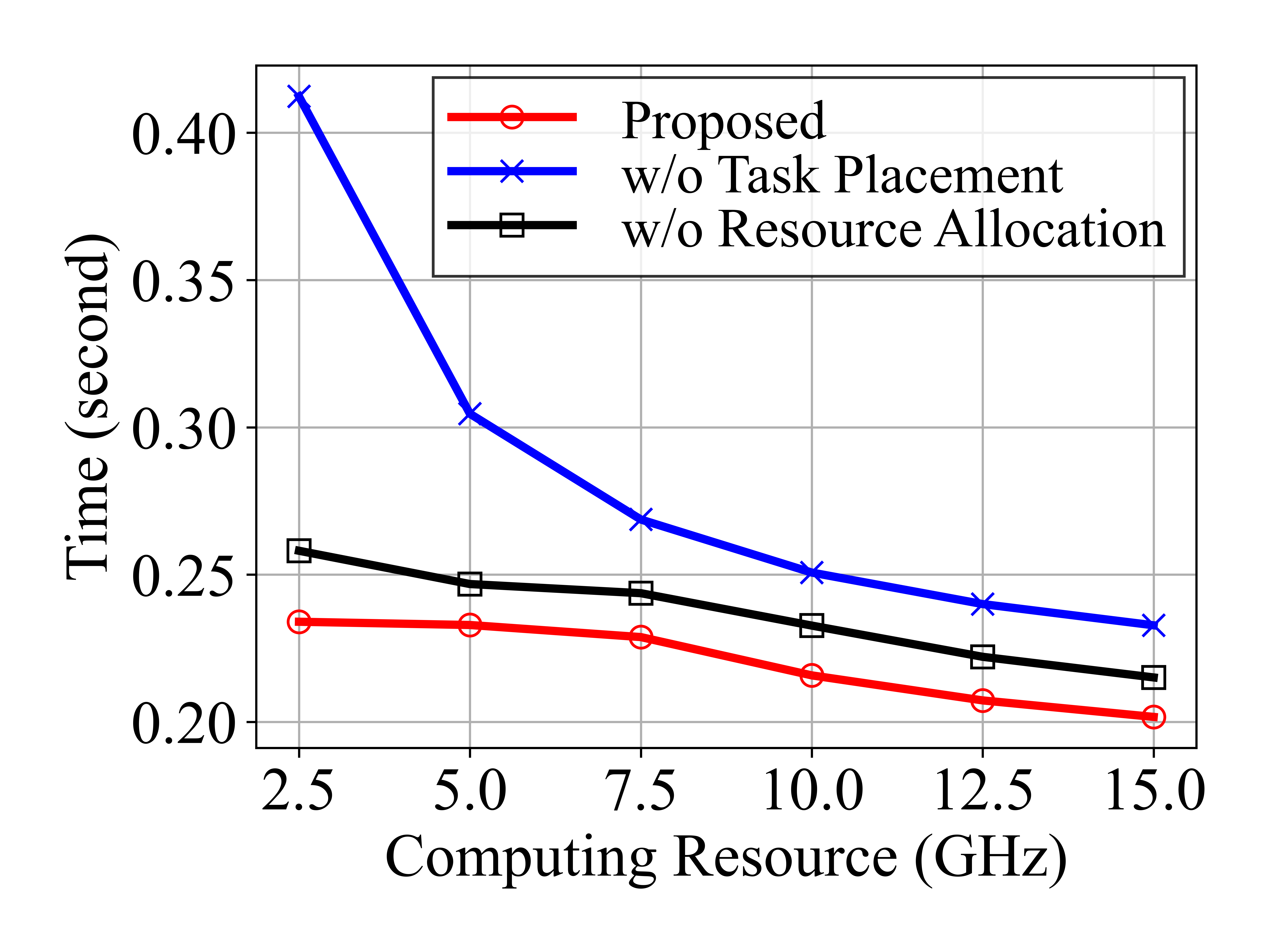}
        \caption{Completion time}
        \label{fig:image3}
    \end{subfigure}
    \vspace{0.5mm}
    \caption{Performance evaluation of the proposed algorithm.}
    \label{fig:three_images}
\end{figure*}

\subsection{Task Placement}
The task placement subproblem can be expressed as $\mathbf{P}4$:
\begin{equation}
\label{P5}
\begin{aligned}
\mathbf{P}4:\ &\min_{\boldsymbol{E}}\sum_{ k \in \mathcal{K}}\sum_{m \in {\mathcal{M}^k}}t_{k,m}\\
&\text{s.t.}\ \eqref{eq:constraint4},\text{and} \ \eqref{P0a}.\\
\end{aligned}
\end{equation}

Problem $\mathbf{P}4$ is a mixed-integer nonlinear programming problem, which cannot be solved by traditional optimization methods. Therefore, we adopt a Gibbs sampling algorithm to solve it. In each iteration, a randomly selected task $\phi_{k,m}$ is virtually modified in terms of its placement decision, and its feasibility is then checked against the constraints. If the constraints are not satisfied, continue to randomly select a placement decision for virtual modification until the constraints are satisfied. Subsequently, $\hat{\boldsymbol{B}}$ and $\hat{\boldsymbol{V}}$ can be calculate based on Algorithm 1 and Equation \eqref{vsolution}. Then, the new objective value $\hat{\Delta}$ is computed, and $\gamma$ is calculated based on $\hat{\Delta}$ and $\Delta$. With probability $\gamma$, the decision variables are updated.
A set of decision variables with a lower time cost is more likely to be accepted. At the end of the iteration, the final decision $\boldsymbol{E}$, $\boldsymbol{B}$, $\boldsymbol{V}$ is returned. Where $\tau$ is the smooth parameter used to balance exploration and exploitation.

\section{Simulation Results}
 
We perform random scenario generation and point cloud data collection in the CARLA simulator and use the point cloud processing algorithm SECOND as the AI model for object classification\cite{SECOND}. We consider a 50-meter-long bidirectional four-lane road segment within the RSU coverage area. The accuracy threshold $A$ is set to 0.85. We created a dataset with 7,350 labels for training the DNN for task accuracy estimation, with 20\% used for testing. The trained DNN achieved an MSE of 0.01. For simplicity, we assume that all CAVs have the same transmission power, channel fading coefficient, and computing resource. The computing resource (GHz) of the CAV is selected among six candidate values in $\{2.5,2.0,7.5,10.0,12.5,15.0\}$, with 10.0 by default. The main simulation parameters are presented in Table I.

The proposed scheme is compared with the following benchmarks:
\begin{itemize}
    \item \textbf{Standalone}: All CAVs independently sense using their own data.
    \item \textbf{Proposed scheme without task placement}: The task placement decision is randomly generated while keeping other aspects consistent with our proposed scheme.
    \item \textbf{Proposed scheme without resource allocation}: The communication and computing resources are evenly distributed while keeping other aspects consistent with our proposed scheme.
\end{itemize}

Figure 2(a) shows the convergence of the proposed algorithm. The objective value is the total task completion time. As can be seen, for $\tau = 10^{-4}$, the algorithm converges to the global optimal solution and it converges the fastest. At this time, if $\tau$ continues to decrease, the convergence speed slows down, like $\tau = 10^{-5}$ and $\tau = 10^{-6}$. The convergence rate is the slowest for $\tau = 10^{-3}$.
Fig. 2(b) shows the sensing accuracy of standalone CAV sensing and our proposed scheme. It can be seen that with the assistance of RSU, the accuracy of each CAV has been improved to a certain extent and has exceeded the required 0.85 threshold. The accuracy of CAV 1 has improved by $38.7 \%$ compared to the original accuracy. 
Fig. 2(c) presents the total task completion time of three schemes when the CAVs have different computing resources. It is observed that our scheme obtains the lowest completion time. Compared to random task placement, our algorithm effectively reduces completion time. The reduction in completion time is significant when CAV computing resources are limited.
\begin{table}[t]
    \centering
    \caption{Simulation Parameters.}
    \footnotesize
    \begin{tabular}{c c|c c}
        \hline\hline
        \textbf{Parameter} & \textbf{Value} & \textbf{Parameter} & \textbf{Value} \\
        \hline
        $B$ & 20 MHz & $B_s$ & 1 MHz \\
        $\xi $& 96 bits & $\omega$ & 50,000 cycles/point \\
        $f_0$ & 200 GHz & $J$ & 3 \\
        $R_1,R_2$ & 20 meters & $L$ & 14 meters \\
        $M$ & 3 & $K$ & 7 \\
        \hline
    \end{tabular}
    \label{tab:simulation_params}
\end{table}

\section{Conclusion}

In this paper, we have designed an RSU-assisted cooperative sensing scheme for CAVs, by efficiently utilizing computing and sensing capabilities provided by the RSU. Additionally, we have designed a two-layer algorithm to judiciously determine task placement, spectrum resource, and computing resource allocation decisions. The proposed scheme can effectively reduce the total completion time of all tasks, as well as satisfy the sensing accuracy requirement of CAVs. For the future work, we will study a mathematical model that can precisely characterize the impact of data quality on sensing accuracy.
\vspace{-0.7\baselineskip}
\bibliographystyle{IEEEtran}
\bibliography{IEEEabrv,ref}

\begin{thebibliography}{10}
\providecommand{\url}[1]{#1}
\csname url@samestyle\endcsname
\providecommand{\newblock}{\relax}
\providecommand{\bibinfo}[2]{#2}
\providecommand{\BIBentrySTDinterwordspacing}{\spaceskip=0pt\relax}
\providecommand{\BIBentryALTinterwordstretchfactor}{4}
\providecommand{\BIBentryALTinterwordspacing}{\spaceskip=\fontdimen2\font plus
\BIBentryALTinterwordstretchfactor\fontdimen3\font minus \fontdimen4\font\relax}
\providecommand{\BIBforeignlanguage}[2]{{%
\expandafter\ifx\csname l@#1\endcsname\relax
\typeout{** WARNING: IEEEtran.bst: No hyphenation pattern has been}%
\typeout{** loaded for the language `#1'. Using the pattern for}%
\typeout{** the default language instead.}%
\else
\language=\csname l@#1\endcsname
\fi
#2}}
\providecommand{\BIBdecl}{\relax}
\BIBdecl

\bibitem{CAVsensor}
Z.~Xiao, J.~Shu, H.~Jiang, G.~Min, H.~Chen, and Z.~Han, ``Perception task offloading with collaborative computation for autonomous driving,'' \emph{IEEE J. Sel. Areas Commun.}, vol.~41, no.~2, pp. 457--473, 2023.

\bibitem{CAVdevelop}
H.~U. Ahmed, Y.~Huang, P.~Lu, and R.~Bridgelall, ``Technology developments and impacts of connected and autonomous vehicles: An overview,'' \emph{Smart Cities}, vol.~5, no.~1, pp. 382--404, Mar. 2022.

\bibitem{CPraw}
Q.~Chen, S.~Tang, Q.~Yang, and S.~Fu, ``Cooper: Cooperative perception for connected autonomous vehicles based on {3D} point clouds,'' in \emph{Proc. IEEE Int. Conf. Distrib. Comput. Syst.}, 2019, pp. 514--524.

\bibitem{CPdecision}
E.~Arnold, M.~Dianati, R.~de~Temple, and S.~Fallah, ``Cooperative perception for {3D} object detection in driving scenarios using infrastructure sensors,'' \emph{IEEE Trans. Intell. Transp. Syst.}, vol.~23, no.~3, pp. 1852--1864, 2022.

\bibitem{rawlevelresearch}
Y.~Jia, R.~Mao, Y.~Sun, S.~Zhou, and Z.~Niu, ``Online {V2X} scheduling for raw-level cooperative perception,'' in \emph{Proc. IEEE Int. Conf. Commun.}, 2022, pp. 309--314.

\bibitem{cooperative}
M.~K. Abdel-Aziz, C.~Perfecto, S.~Samarakoon, M.~Bennis, and W.~Saad, ``Vehicular cooperative perception through action branching and federated reinforcement learning,'' \emph{IEEE Trans. Commun.}, vol.~70, no.~2, pp. 891--903, 2022.

\bibitem{dataquality}
X.~Ye, K.~Qu, W.~Zhuang, and X.~Shen, ``Accuracy-aware cooperative sensing and computing for connected autonomous vehicles,'' \emph{IEEE Trans. Mobile Comput.}, vol.~23, no.~8, pp. 8193--8207, 2024.

\bibitem{rawlevelresearch2}
G.~Luo, C.~Shao, N.~Cheng, H.~Zhou, H.~Zhang, Q.~Yuan, and J.~Li, ``Edgecooper: Network-aware cooperative {LiDAR} perception for enhanced vehicular awareness,'' \emph{IEEE J. Sel. Areas Commun.}, vol.~42, no.~1, pp. 207--222, 2024.

\bibitem{RSUresource1}
X.~Shen, J.~Gao, W.~Wu, M.~Li, C.~Zhou, and W.~Zhuang, ``Holistic network virtualization and pervasive network intelligence for {6G},'' \emph{IEEE Commun. Surveys Tuts.}, vol.~24, no.~1, pp. 1--30, 2022.

\bibitem{RSUsensor}
Y.~Zhang, N.~Bhattarai, J.~Zhao, H.~Liu, and H.~Xu, ``An unsupervised clustering method for processing roadside {LiDAR} data with improved computational efficiency,'' \emph{IEEE Sensors J.}, vol.~22, no.~11, pp. 10\,684--10\,691, 2022.

\bibitem{RoI}
Z.~Xiao, J.~Shu, H.~Jiang, G.~Min, J.~Liang, and A.~Iyengar, ``Toward collaborative occlusion-free perception in connected autonomous vehicles,'' \emph{IEEE Trans. Mobile Comput.}, vol.~23, no.~5, pp. 4918--4929, 2024.

\bibitem{RSUvehicleresource}
W.~Fan, Y.~Su, J.~Liu, S.~Li, W.~Huang, F.~Wu, and Y.~Liu, ``Joint task offloading and resource allocation for vehicular edge computing based on {V2I} and {V2V} modes,'' \emph{IEEE Trans. Intell. Transp. Syst.}, vol.~24, no.~4, pp. 4277--4292, 2023.

\bibitem{SECOND}
Y.~Yan, Y.~Mao, and B.~Li, ``Second: Sparsely embedded convolutional detection,'' \emph{Sensors}, vol.~18, no.~10, pp. 1--17, 2018.

\bibitem{RSUresource2}
K.~Qu, W.~Zhuang, Q.~Ye, W.~Wu, and X.~Shen, ``Model-assisted learning for adaptive cooperative perception of connected autonomous vehicles,'' \emph{IEEE Trans. Wireless Commun.}, vol.~23, no.~8, pp. 8820--8835, 2024.

\end{thebibliography}

\end{document}